\documentclass[fleqn,12pt,twoside]{article}
\usepackage{espcrc1}

\usepackage{graphicx}

\newcommand{\AmS}{{\protect\the\textfont2
  A\kern-.1667em\lower.5ex\hbox{M}\kern-.125emS}}

\title{Observations of r- and s-process elements in Population II stars}

\author{Sean G. Ryan\address{Department of Physics and Astronomy,
	The Open University, Walton Hall, Milton Keynes MK7 6AA,
	United Kingdom. s.g.ryan@open.ac.uk}
	\thanks{Nuclei in the Cosmos VII, Fuji-Yoshida, 8-12 July 2002}
	}
       
\begin{document}

\maketitle

\begin{abstract}
The framework for the interpretation of neutron-capture elements observed
in Population II stars, established 20-25 years ago, is that these stars primarily
exhibit r-process signatures, due to the inefficiency of the s-process at
low metallicity. A view developed later that the r-process might be
universal, which is to say that the same r-process element ratios would
exist at high and low metallicity. However, observations of s-process
abundances in low-metallicity environments, and departures
from a universal r-process ratio for the lightest neutron-capture elements
have required revisions to the framework.
Observations indicating the need for and nature of these changes will be
presented.
\end{abstract}

\section{Introduction}

Elements beyond the iron peak are formed in neutron-capture reactions.
Traditionally these are split into two categories, the slow- (s-) process
when beta-decay rates exceed neutron-capture rates, and the rapid- (r-) process
when neutron-capture rates are high enough to produce neutron-rich unstable
isotopes far from the valley of beta-stability.

The main site for the r-process is currently thought to be supernovae,
though neutron-star collisions may have played a role in later stages
of Galactic enrichment \cite{a02}. The s-process has contributions from two
sites: thermally-pulsing (TP-) AGB stars are responsible for the 
so-called `main'
s-process fraction, while an additional contribution from high-mass 
He-core-burning stars is required to explain the first neutron-capture peak 
(nuclei with closed neutron shells having a neutron number $N$ = 50; Sr, Y, Zr).
A third s-process, the so-called 
`strong' s-process, that was invoked to explain the third neutron-capture
peak (closed-neutron-shell nuclei with $N$ = 126; Pb and Bi), appears now
to be associated with metal-poor AGB stars \cite{gabltscl98}. 

In metal-poor environments, such as existed during the formation of the oldest 
stars in the Galaxy, the s-process has long been regarded as unimportant.
This interpretation was established some 20-25 years ago. 
The observational work of 
Spite and Spite\cite{ss78} showed that the abundance of Ba dropped
faster than the abundance of Eu as progressively more-metal-poor stars were
observed. As Eu is an almost pure r-process element, whereas Ba can have
contributions from both processes, this was interpreted as evidence for the
disappearance of the s-process contribution to Ba in metal-poor stars.
The theoretical basis for the framework was provided by 
Truran\cite{t81} who pointed out that the lack of seed nuclei 
would block the s-process in the metal-poor counterparts of the objects which,
at solar-metallicity, are primarily responsible for the s-process.
The r-process, on the other hand, would be capable of generating its
own seed nuclei even in metal-poor environments.

This basic framework was supported by various observations and modelling
efforts in subsequent years. 
Gilroy et al. \cite{gspc88} showed that in metal-poor
giants, neutron-capture elements have the same relative abundances as in the 
inferred r-process contribution to the solar composition. 
Furthermore, detailed calculations of Galactic chemical 
evolution \cite{tggbfs99} show that AGB stars would
not contribute significantly to Galactic enrichment of s-process elements until
[Fe/H] had climbed to $\sim$ $-1$.

Observations of [Ba/Eu] ratios were extended to 
[Fe/H] $\simeq$ $-3$ and showed that the solar r-process ratio persisted
\cite{mpss95}. The star CS~22892-052, with [Fe/H] $\simeq$ $-2.5$ and
having abundance ratios of r-process to iron-peak elements
forty times higher than normal,
likewise had a composition
in which r-process abundances were seen \cite{spms94,smpcba96}.
The relative weakness of iron-peak elements meant that a vast array of 
neutron-capture elements could be studied, 
rather than just the more easily detected ones such as Ba and Eu, and hence the 
existence of an r-process pattern was established far more thoroughly for this
star than for any other metal-poor object. This included filling in the gaps
in the coverage of elements between the neutron capture peaks
\cite{cstb96,scbt98,cskbd98}.

A view developed that the r-process might be ``universal'', i.e. 
that the element yields in the r-process were independent of the metallicity 
of the gas from which the star providing the site of nucleosynthesis formed.
This view was supported by findings that the element ratios near the second
neutron-capture peak (closed-neutron-shell nuclei with $N$ = 126; e.g. Ba, La)
resembled the solar r-process even at low [Fe/H] 
\cite{gspc88,cbsmp95,cstb96,smpcba96}.
As we shall see below, the concept of universality is still entertained, but
observations that Sr exhibits a very different behaviour
\cite{rnb91,npb93,mpss95,rnb96} have seen the idea of universality
qualified with the restriction that is does not apply to elements in
the first neutron-capture peak.

CS~22892-052 soon became, and probably rightly so,
one of the most talked about stars in the field of Pop. II
nucleosynthesis. Less justifiably, it was often spoken of
as an archetypal metal-poor star for the purposes of investigating
neutron-capture nucleosynthesis, an epithet which overlooked two 
key observational results: 
(1) its fame stemmed from it being {\it abnormally} rich in neutron-capture 
elements, very definitely {\it not} representative of the rest of 
Pop. II, and
(2) it was one of a growing class of carbon-rich, metal-poor stars.

Carbon-rich, metal-poor stars were found \cite{bps85,bps92} in follow-up
observations of stars originally identified as metal-poor on
objective-prism spectra of the Ca H and K and H$\epsilon$ lines.
The objective prism spectra covered none of the CH or CN molecular features,
so contained no information on carbon abundances. However, blue,
$\sim$1~\AA-resolution spectra obtained subsequently
showed that a significant fraction of
such objects contained unusually strong CH bands; 
up to 25\%\ of very metal-poor stars were suspected of having significant
carbon excesses. The excesses were eventually shown \cite{rbs99} to extend up to
two orders of magnitude, giving [C/Fe] = 2.0.

In work aiming to clarify the relationship between the C and r-process 
enhancements in CS~22892-052, and with the hope of identifying more
stars whose r-process elements could be studied in detail, 
Norris, Beers and Ryan set out to observe a number of C-rich stars identified
from 1~\AA-resolution spectra of objective-prism-selected metal-poor objects.
Their early observations \cite{nrb97a}, and extended studies involving
many of the other targets \cite{arnbaikmf02,anrba02}, showed not a
single additional C-rich star with an r-process enhancement, but
revealed instead large numbers of metal-poor, C-rich stars with clear
s-process signatures (see also \cite{bcsbsnn97,hbsscpbnan00}), as well as
others with no neutron-capture anomalies \cite{nrb97b}.

In the remainder of this paper, we discuss two challenges to the 
standard framework for understanding neutron-capture nucleosynthesis in
Pop. II stars: 
the existence of a significant number of Pop.
II stars with enhanced carbon and s-process elements, and
the apparent independence of Sr from the 
behaviour of Ba.

\section{C-rich, metal-poor stars}

The abundant s-process elements found in many of the C-rich stars
were unexpected given the strong arguments (presented above) that 
favoured an
r-process origin for the neutron-capture species in metal-poor stars.
High abundances of C and s-process elements initially suggested that
AGB stars may have been active, since AGB stars are the major sites of 
both C and s-process element production in the Galaxy. However,
two of the first few C-rich, metal-poor stars examined in detail were 
subgiants rather than giants, having being found originally in a study of
subdwarfs selected on the basis of high-proper motion \cite{rn91}, so it
was shown quickly that the stars being observed with high C and s-process 
abundances could not themselves be responsible for the enrichment. Since
the observed stars could not be AGB stars, and the contribution of
AGB stars to Galactic chemical enrichment is believed be insignificant at
[Fe/H] $\sim$ $-2.5$, the search turned for means of directly contaminating
them with the products of AGB nucleosynthesis {\it after} their formation.
Suspected mechanisms therefore centred on the accretion of
mass lost from an AGB companion, either via Roche-lobe overflow or perhaps
more commonly via wind accretion. This mechanism matches the binary mechanism
responsible for the origin of Pop. II CH stars and Pop. I Ba stars
\cite{mfn80,lb91,hetp95,v92}.

If this mechanism were to be able to explain the origins of these stars,
we would require binary periods long enough (i.e. separations large enough)
to avoid mass transfer on the RGB instead of the AGB, and yet short enough for 
wind accretion to be efficient. These factors imply periods in the range
1-10~yr \cite{jb92,hetp95,w97}.  We would expect to see the chemical signature
of a low-metallicity AGB star, and radial velocity variations on the stated
timescale due to the presence of a white dwarf companion, being the compact 
remnant of the AGB star.

A mechanism proposed for the CH stars, before their
binarity and its role was discovered,
was that the He flash that terminates RGB evolution, or a He shell 
flash on the TP-AGB, triggered a deep mixing event that refuelled the core and
re-ignited core H burning, 
hence providing it with a second life \cite{b74}.
Although this possibility fell out of favour, the idea has been revived in
a slightly different context, and mixing events triggered
by the He flash in zero-metallicity (Pop. III) stars have been investigated
more recently \cite {fii00,scsw01}. They may provide an alternative means of
producing C-rich, metal-poor objects.

Observations of s-process enhanced stars have provided mixed support for
the binary-star, mass-transfer hypothesis. LP~625-44, which has [Fe/H] = $-2.7$,
[C/Fe] = 2.0 and [Ba/Fe] = 2.6, has been found to have a low-luminosity
companion with an orbital period $P$ $^>_\sim$ 13~yr (\cite{anrba00}; see also
Figure~\ref{fig:radvel} below). Furthermore,
the s-process abundance ratios in this object from Ba to Pb
are consistent with 
production in a low-metallicity AGB star (where the efficiency of the
s-process is treated as a free parameter) \cite{ranbgba01}.
The observations suggest a low mass for the AGB progenitor.

However, LP~706-7, which has almost identical abundance ratios to LP~625-44
([Fe/H] = $-2.7$, [C/Fe] = 2.0 and [Ba/Fe] = 2.0) shows no 
significant radial-velocity variations over 11 years of observation; see
Figure~\ref{fig:radvel}. LP~706-7 has an effective temperature of 
6250~K and lies on the subgiant branch between the main-sequence turnoff and the
base of the giant branch. Like most other Pop.~II stars in this phase of
evolution, it retains the primordial Li abundance. 
This last point is very difficult to reconcile with either a mass-transfer 
or mixing scenario for the origin of its C and s-process excesses.
The reason is that Li is extremely fragile, and survives only in the outermost
few percent (by mass) of the star. Any process which involves mixing
of the surface with interior layers, or requires that the surface layers
originated in a more evolved star, would fail to account for the existence
or primordial abundance of this element. Although Li can be produced in
evolved stars, producing it at just the right level to match the
primordial value would be a severe challenge, and hence is unlikely to
explain this case.

\begin{figure}[htb]
\includegraphics[scale=0.85]{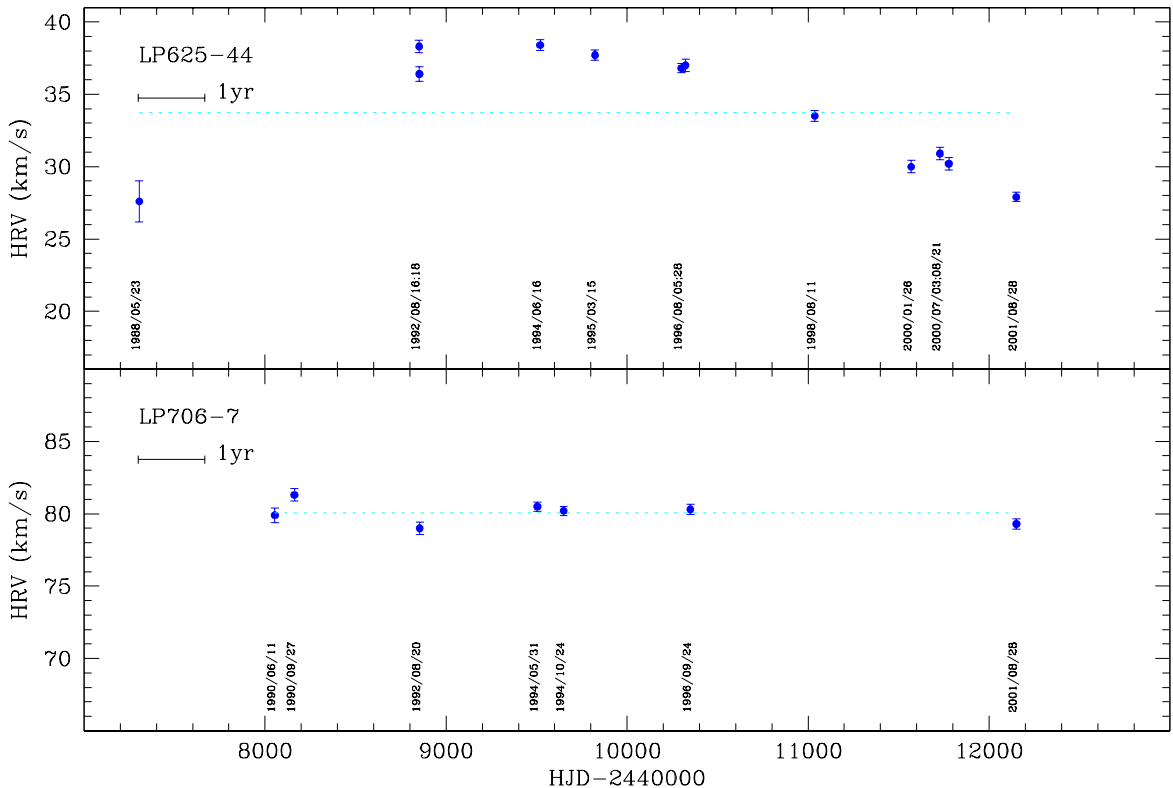}
\caption{Radial velocity variations of LP~625-44 and LP~706-7 over 
$\simeq$ 10~yr. Both stars are C rich and s-process rich, 
and have similar chemical compositions, but LP~706-7 exhibits no 
significant radial velocity
variations and also has a Li/H ratio at the primordial value.
The horizontal dashed line gives the mean of the observations.
Error bars are 1$\sigma$.}
\label{fig:radvel}
\end{figure}

Observations of other C-rich, metal-poor stars close to the main-sequence 
turnoff \cite{ps01} likewise show a significant lack of radial-velocity 
variations.
This, combined with the problem of Li survival in LP~706-7, emphasises
that while 
the AGB mass-transfer explanation for the C-rich, s-process-rich stars
may be attractive in that it fits with established phenomena in the evolution
of stars, clearly it is not the answer for all of the s-process-rich,
C-rich Pop.~II stars. Even if AGB stars can produce the element ratios
seen in these objects, the problem of how it is incorporated into the
atmospheres of these objects is not yet resolved.

\section{Neutron-capture elements in C-normal, very metal-poor stars}

As discussed in the introduction, wide variations (up to a factor
of 100) in the [Sr/Fe] and [Sr/Ba] ratios of stars with [Fe/H] $<$ $-3$
have been observed \cite{rnb91,npb93,mpss95,rnb96}.
Star-to-star variations in [$r$/Fe] have also been seen for a range of
other r-process elements, e.g. \cite{gspc88}, but these seem to only rarely
reach large extremes, whereas Sr regularly exhibits large variations.
It was the decoupling of the behaviour of Sr from that of Ba which first
indicated that a universal r-process could not exist. Current evidence
is consistent with a universal r-process for stable elements at the
second neutron-capture peak and above, but clearly not for the first
neutron-capture peak. At present there are too few observations of elements 
between the first and second peaks in stars with [Fe/H] $<$ $-3$ to know
at which atomic number universality is established. That is, we do not know
whether silver correlates with Sr, with Ba, or with neither in the most
metal-poor stars.

The lack of data on Ag may be expected since observations of the lines
of this element, which are in the UV,  are extremely rare.
Europium, on the other hand, is frequently measured in the blue spectra of
Pop.~II stars, but the lines become very weak at [Fe/H] $<$ $-3$. 
Unfortunately, that is precisely where the Sr anomalies become most
evident. The result of this lack of data on Eu in very metal poor stars
is that we do not usually have measurements much above the second r-process
peak in the most metal-poor stars. As Eu is almost a pure r-process element,
it avoids any ambiguity associated with the s- and r-process fractions
of Ba.
The origin of Ba was called into question when analysis of its
hyperfine structure suggested an odd-to-even isotope ratio more in keeping
with the s-process than the r-process \cite{m95}, though this result has 
been challenged by new observations \cite{la02}.
More importantly, Eu probes a different region of the chart of nuclides
than Ba. 

The lack of observations of Eu has until very recently \cite{f00,jb01}
meant that the region with [Fe/H] $<$ $-3$ was devoid of Eu detections,
constrained only by quite high upper limits on [Eu/Fe]. In addition to
the new data cited, observations with the new HDS spectrograph on the Subaru
telescope have permitted an extension of the observations down towards 
[Fe/H] = $-3.4$. These observations \cite{iwra02} confirm that [Eu/Fe] is only
slightly sub-solar, $\simeq$ $-0.2$. This, combined 
with the previously known [Ba/Fe] decline at low metallicity,
indicates that [Ba/Eu] is close to the solar r-process fraction in these 
objects. In contrast, [Sr/Eu] remains above the solar r-process fraction.

This finding, combined with the high and unpredictable
[Sr/Fe] values in such metal-poor stars,
adds weight to the suggestion that the r-process sets a lower limit on the
production of Sr, which one may infer from the [Sr/Ba] ratio that would
be found if the r-process contribution were truly universal even for
elements at the first neutron-capture peak. On top of this, an additional
Sr source is required that appears to be
highly variable from one nucleosynthesis
site to the next \cite{rnb98}.

Possible mechanisms \cite{r00} are split between the known and the unknown.
The former category includes the weak s-process, which is known to produce
only first neutron-capture-peak elements, but which requires pre-existing 
metals both as the $^{22}$Ne neutron source and for the
neutron-capture seed nuclei, and hence is expected to be ineffective
in very metal-poor stars. The alpha-process (alpha-rich freeze-out)
may be an alternative if nucleosynthesis can extend to Sr in
significant numbers. Unknown processes that may mimic the weak s-process
yields include a weak-r-process \cite{iw99}, and possibly a 
neutron-capture process that resembles neither of the 
traditional extremes (s- and r-) of neutron-capture timescales.

An element that may help resolve the origin of the Sr excesses is zinc.
At atomic number $Z_{\rm Zn}$ = 30, this lies between the iron peak 
($Z_{\rm Fe}$ = 26) and Sr ($Z_{\rm Sr}$ = 38). Contributions to its
nucleosynthesis come from both
the alpha-process and the weak-s-process, and hence it may be able to 
indicate
whether one or other of these processes has been active.

\begin{figure}[htb]
\includegraphics[scale=0.85]{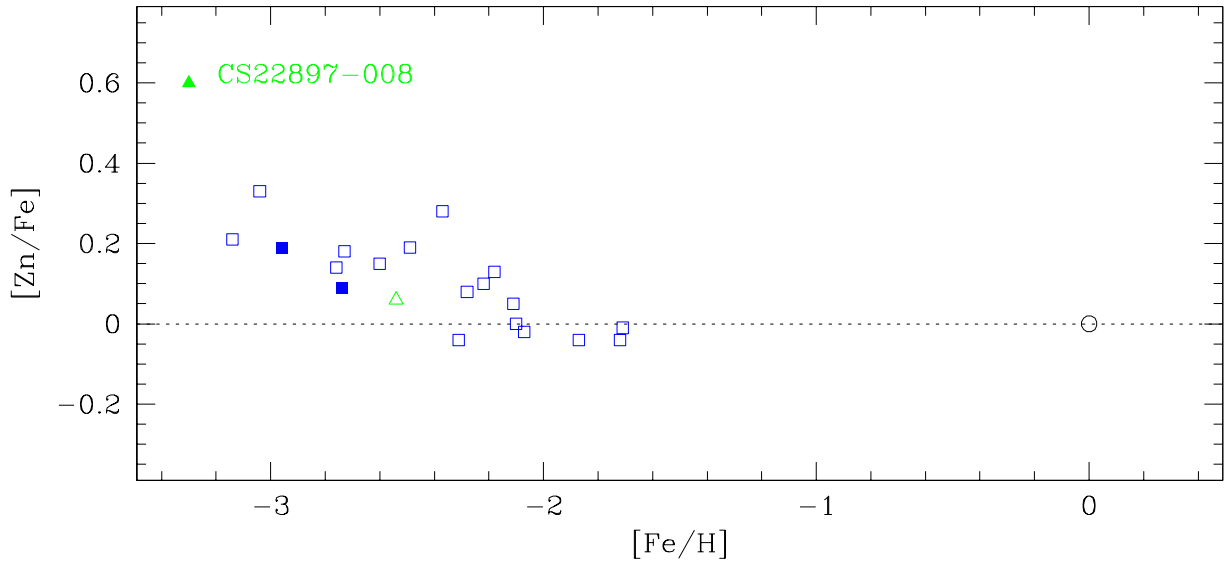}
\caption{Zn and Fe abundances for 
normal giants (open squares \cite{jb02a}),
Y-rich giants (filled squares \cite{jb02b}),
HD~140283 (open triangle \cite{b03}), 
and
the Sr-rich giant CS~22897-008 (filled triangle \cite{brnb01}).
}
\label{fig:znfe}
\end{figure}

The possible link between Zn and Sr was first tested with the
star CS~22987-008 \cite{r00b,brnb01}, which was previously 
discovered to be very Sr-rich 
\cite{mpss95}. 
The star was indeed found to have [Zn/Fe] above solar, but around the same time
evidence emerged that so may large numbers of very metal-poor stars
\cite{pbskbbd00,jb01b}. 
The comparison in Figure~\ref{fig:znfe} 
shows data for 
CS~22897-008 \cite{brnb01} and the normal star HD~140283 \cite{b03}
against metal-poor giants investigated
by Johnson and Bolte \cite{jb02a,jb02b}. Although \cite{jb02b} could not
calculate Sr abundances, they found high [Y/Ba] values in several of 
their objects. One might expect these same objects to have
high [Sr/Ba] values; this could be tested.
The high-Y stars have been highlighted in the figure. 
Although the genuinely Sr-rich
giant, CS~22897-008, has a [Zn/Fe] value even higher than the 
rising trend, the high-[Y/Ba] stars do not. If high-Sr and high-Y stars
do belong together, then the lack of distinction between the Y-rich and
Y-normal stars in the [Zn/Fe] vs [Fe/H] diagram may indicate that the
high Zn abundance of CS~22897-008 is not related to its high Sr.
Analysis of these objects is continuing, but it appears that the
interpretation of the observations will not be simple!
The nucleosynthesis of elements above the iron peak retains at least some 
of its secrets 20 years after the framework for understanding metal-poor
stars was established.

\section{Acknowledgments}

This paper reflects collaborations and discussions with 
numerous scientists, especially W. Aoki, T. C. Beers, L. A. J. Blake,
R. Gallino, Y. Ishimaru,
J. E. Norris, S. Tsangarides, and S. Wanajo, to whom I am grateful
for their contributions. I am also very
gratefully to the organising committee for
financial assistance that made my attendance possible.

\end{document}